\pdfoutput=1
\RequirePackage{ifpdf}
\ifpdf 
\documentclass[pdftex]{sigma}
\else
\documentclass{sigma}
\fi

\numberwithin{equation}{section}

\begin{document}

\allowdisplaybreaks

\renewcommand{\PaperNumber}{084}

\FirstPageHeading

\ShortArticleName{Quasi-Grammian Solutions of the Generalized Coupled Dispersionless Integrable System}

\ArticleName{Quasi-Grammian Solutions of the Generalized\\ Coupled Dispersionless Integrable System}

\Author{Bushra HAIDER and Mahmood-ul HASSAN}

\AuthorNameForHeading{B.~Haider and M.~Hassan}

\Address{Department of Physics, University of the Punjab,\\
Quaid-e-Azam Campus, Lahore-54590, Pakistan}
\Email{\href{mailto:bushrahaider@hotmail.com}{bushrahaider@hotmail.com},
\href{mailto:mhassan.physics@pu.edu.pk}{mhassan.physics@pu.edu.pk}}
\URLaddress{\url{http://pu.edu.pk/faculty/description/526/},\\
\hspace*{10.5mm}\url{http://www.pu.edu.pk/faculty/description/538/}}

\ArticleDates{Received June 22, 2012, in f\/inal form October 10, 2012; Published online November 08, 2012}

\Abstract{The standard binary Darboux transformation is investigated
and is used to obtain quasi-Grammian multisoliton solutions of the
generalized coupled dispersionless integrable system.}


\Classification{70H06; 22E99}

\section{Introduction}

The interest in dispersionless integrable systems is due to their
wide range
of applicability in various f\/ields of mathematics and physics~\cite{Aoyama,Caroll,M.Dunajski,Mhassan,hirota,kaku1,kaku, Kakuhata:1995ye, kodama,kodam,konno,konopel,kot,Krichever,Takasaki1,Takasaki,Wiegmann}. Most of the dispersionless integrable systems belong
to a family where these systems arise as quasi-classical limit of
ordinary integrable systems with a dispersion term \cite{Aoyama,Caroll,M.Dunajski,kodama,kodam,konopel,Krichever,Takasaki1,Takasaki,Wiegmann}. But there are important examples of dispersionless
integrable system which are referred to as dispersionless not in the
sense mentioned above but due to the absence of dispersion term. The
coupled dispersionless integrable systems
and its generalizations are examples of such integrable systems \cite{Mhassan, hirota,kaku1,kaku, Kakuhata:1995ye,konno,kot}. The Darboux transformation of the generalized
coupled
dispersionless integrable system has been studied in a~recent work~\cite{Mhassan}.

The purpose of this paper is to study the standard binary Darboux
transformation of the generalized coupled dispersionless integrable
system and to derive exact solutions in terms of quasi-Grammians. We
employ the method introduced in~\cite{BDTPCM}, construct standard
binary Darboux transformation by introducing Darboux matrices of the
system for the direct and the adjoint Lax pairs and then obtain
binary Darboux matrix by composing the two Darboux transformations.
We obtain the quasi-Grammian multisoliton solutions using the
iterated binary Darboux transformations. We also consider the system
based on Lie group $SU(N) $ and obtain explicit
solutions of the system based on $SU(2) $.

The action of the generalized coupled dispersionless integrable
system based
on some non-Abelian Lie group $\mathcal{G}$ is given by%
\begin{gather}
I=\int dtdx\mathcal{L}( S,S_{x},S_{t}) ,  \label{action}
\end{gather}%
where the Lagrangian density $\mathcal{L}( S,S_{x},S_{t})
$ is
def\/ined by
\begin{gather*}
\mathcal{L}=\operatorname{Tr}\left( \frac{1}{2}S_{x}S_{t}-\frac{1}{3}G[ S,[ S_{x},S] ] \right) , 
\end{gather*}
where $S$ is a matrix f\/ield and $G$ is a constant matrix taking
values in the non-Abelian Lie algebra $\mathfrak{g}$ of the Lie
group~$\mathcal{G}$.
The matrix f\/ields $S$ and~$G$ are Lie algebra $\mathfrak{g}$ valued, i.e., $S=\phi ^{a}T^{a}$ and $G=\kappa ^{a}T^{a}$, where anti-hermitian generators $ \{ T^{a},\; a=1,2,\ldots ,\dim\mathfrak{g} \} $ of
the Lie algebra~$\mathfrak{g}$ obey $ [ T^{a},T^{b} ]
=f^{abc}T^{c}$ and $\operatorname{Tr} ( T^{a}T^{b} ) =-\delta
^{ab}$. For any $X\in \mathfrak{g}$, $X=X^{a}T^{a}$. Note that $\phi
^{a}=\phi ^{a}\left( x,t\right) $ is a vector f\/ield with components
$ \{ \phi ^{q},\; a=1,2,\ldots ,\dim \mathfrak{g} \} $
and $\kappa $ is the
constant vector having components $ \{ \kappa ^{a},\;a=1,2,\ldots ,\dim \mathfrak{g} \} .$ The equation of motion of the
generalized
coupled dispersionless system as obtained from (\ref{action}) is
\begin{gather}
S_{xt}- [  [ S,G ] ,S_{x} ] =0.  \label{EQOM}
\end{gather}
For $\mathcal{G}=SU ( 2 ) $ we get from (\ref{EQOM})
\begin{gather}
q_{xt}+ ( r\bar{r} ) _{x}  = 0,  \qquad
r_{xt}-2q_{x}r  = 0,  \qquad
\bar{r}_{xt}-2q_{x}\bar{r}  = 0,  \label{SU2}
\end{gather}
where $q$ is a real valued function and $r$ is a complex valued function of $%
x$ and~$t$. Here $\bar{r}$ denotes complex conjugate of~$r$.

The generalized coupled dispersionless system (\ref{EQOM}) can be
written as the compatibility condition of the following Lax pair
\begin{gather}
 {\partial }_{x}\psi  = U ( x,t,\lambda  ) \psi ,  \qquad
\partial _{t}\psi  = V ( x,t,\lambda  ) \psi ,    \label{LAXD}
\end{gather}
where $\psi \in \mathcal{G}$ and ${\lambda }$ is a real (or complex)
parameter. The f\/ields~$U$ and~$V$ are $n\times n$ matrix f\/ields and
are
given by
\begin{gather*}
U ( x,t,\lambda  )  = \lambda \partial _{x}S, \qquad 
V ( x,t,\lambda  )  =  [ S,G ] +\lambda ^{-1}G.
\end{gather*}
The compatibility condition of the linear system (\ref{LAXD}) is the
zero
curvature condition
\begin{gather}
\partial _{t}U ( x,t,\lambda  ) -\partial _{x}V ( x,t,\lambda
 ) + [ U ( x,t,\lambda  ) ,V ( x,t,\lambda  )
 ] =0.  \label{ZCD}
\end{gather}
Note that the above equation (\ref{ZCD}) is equivalent to the
equation of motion (\ref{EQOM}). The Darboux transformation of the
generalized coupled dispersionless system has been discussed in~\cite{Mhassan}. In the next section we will retrace the steps for
the Darboux transformation for direct and adjoint spaces and then we
will combine the two elementary Darboux transformations to obtain
the standard binary Darboux transformation of the generalized
coupled dispersionless system.

\section{Darboux transformation on the direct and adjoint Lax
pairs}
In this section we discuss the Darboux transformation on the
solutions to the direct and adjoint Lax pairs. For details of
Darboux transformation see e.g.~\cite{cie,cie1,darboux1,darboux,Leble3,Hassan2008,qing_ji,Leble2,Leble1,Leble+,Leble,man,matveev4,matveev1,
matveev2,matveev3,matveev,park,sakh,Ustinov:1998ga}. The
one-fold Darboux transformation on the matrix solution to the Lax
pair (\ref{LAXD}) is def\/ined by
\begin{gather}
\tilde{\psi}(\lambda )=D(x^{+},x^{-},{\lambda })\psi (\lambda )  ,
\label{Vd}
\end{gather}
where $D(x,t,{\lambda })$ is the Darboux matrix. We use the
following ansatz for the Darboux matrix $D(x,t,{\lambda })$
\begin{gather}
D(x^{+},x^{-},\lambda )=\lambda ^{-1}I-M(x^{+},x^{-}),
\label{darbouxm1}
\end{gather}%
and $M(x^{+},x^{-})$ is some $n\times n$ matrix f\/ield to be determined and $%
I $ is an $n\times n$ identity matrix. The Darboux matrix transforms
the
matrix solution $\psi $ in space $\mathcal{V}$ to a new solution $\tilde{\psi%
}$ in $\widetilde{\mathcal{V}}$, i.e.
\begin{gather}
D({\lambda }): \ \mathcal{V}\rightarrow \widetilde{\mathcal{V}}.
\label{directtrans}
\end{gather}%
The new solution $\tilde{\psi}$ satisf\/ies the Darboux transformed
Lax pair
\begin{gather}
  {\partial }_{x}\tilde{\psi}  = \tilde{U} ( x,t,\lambda  ) \tilde{\psi},  \qquad
\partial _{t}\tilde{\psi}  = \tilde{V} ( x,t,\lambda  ) \tilde{\psi},   \label{LAXD1}
\end{gather}
where the matrix-valued f\/ields $\tilde{U}$ and $\tilde{V}$ are given as%
\begin{gather*}
\tilde{U} ( x,t,\lambda  )  = \lambda \partial
_{x}\tilde{S},\qquad
\tilde{V} ( x,t,\lambda  )  = \big[
\tilde{S},\tilde{G}\big] +\lambda ^{-1}\tilde{G},  
\end{gather*}
and $\tilde{S}$ and $\tilde{G}$ are the Lie algebra valued
transformed matrix f\/ields. The covariance of the Lax pair
(\ref{LAXD}) under Darboux transformation can be checked by
substituting equation~(\ref{Vd}) in equations~(\ref{LAXD1}). The
covariance implies the following Darboux transformation on the
matrix-valued f\/ields~$S$ and~$G$
\begin{gather}
\tilde{S}  = S-M,  \label{S1} \\
\tilde{G}  = G.  \label{G1}
\end{gather}
As mentioned earlier equation (\ref{G1}) shows that $G$ is a
constant matrix and the matrix $M$ is subjected to satisfy the
following equations
\begin{gather*}
\partial _{x}MM  =  [ \partial _{x}S,M ] , \qquad
\partial _{t}M  =  [  [ S,G ] ,M ] + [ G,M ] M.
\end{gather*}
The matrix $M$ can be written in terms of the solutions of the
linear system \cite{Mhassan}
\begin{gather}
M=\Theta \Lambda ^{-1}\Theta ^{-1},  \label{SM1}
\end{gather}
where $\Theta $ is the particular matrix solution of the Lax pair def\/ined by
\begin{gather*}
\Theta =( \psi ({\lambda }_{1})\vert 1\rangle ,\dots ,\psi ({\lambda }_{n})\vert n\rangle ) =( \vert
\theta _{1}\rangle ,\dots ,\vert \theta
_{n}\rangle ) ,
\end{gather*}
Each column $ \vert \theta _{i} \rangle =\psi ({\lambda }%
_{i}) \vert i \rangle $ in $\Theta $ is a column solution
of the Lax pair (\ref{LAXD}) when ${\lambda }={\lambda }_{i}$, i.e.,
it satisf\/ies
\begin{gather}
\ {\partial }_{x} \vert \theta _{i} \rangle  = \lambda
_{i}\partial
_{x}S \vert \theta _{i} \rangle ,  \qquad
\partial _{t} \vert \theta _{i} \rangle  =  [ S,G ]
 \vert \theta _{i} \rangle +\lambda _{i}^{-1}G \vert
\theta _{i} \rangle ,    \label{THETA}
\end{gather}
and $i=1,2,\dots ,n$. Assuming $\Lambda =\operatorname{diag}({\lambda }_{1},\dots ,{\lambda }_{n}),$ the equations (\ref{THETA}) can be written in
matrix form
as
\begin{gather*}
\ {\partial }_{x}\Theta =\partial _{x}S\Theta \Lambda ,  \qquad
\partial _{t}\Theta  =  [ S,G ] \Theta +G\Theta \Lambda ^{-1}.
\end{gather*}
The Darboux transformation of the generalized coupled dispersionless
integrable system in terms of particular matrix solution $\Theta $
with the
particular eigenvalue matrix $\Lambda $ is given as%
\begin{gather*}
\tilde{\psi}  = \left( \lambda ^{-1}I-\Theta \Lambda ^{-1}\Theta
^{-1}\right) \psi ,  \qquad
\tilde{S}  = S-\Theta \Lambda ^{-1}\Theta ^{-1},  \qquad
\tilde{G}  = G.  
\end{gather*}
In terms of quasideterminants we can write the above expressions as%
\begin{gather*}
\tilde{\psi}  = \left\vert
\begin{array}{cc}
\Theta & \psi \\
\Theta \Lambda ^{-1} & \text{\fbox{$\lambda ^{-1}\psi $}}
\end{array}%
\right\vert , \qquad 
\tilde{S}  = S+\left\vert
\begin{array}{cc}
\Theta & I \\
\Theta \Lambda ^{-1} & \text{\fbox{$O$}}
\end{array}
\right\vert ,  
\end{gather*}
where $O$ is an $n\times n$ null matrix. The result can be
generalized to obtain $K$-fold Darboux transformation on matrix
solution $\psi $ and can be
written in terms of quasideterminant as\footnote{The quasideterminant for an $N\times N$ matrix over a ring $R$ is
def\/ined as
\begin{gather*}
|X|_{ij}=\left\vert
\begin{array}{cc}
X^{ij} & c_{j}^{\,\,i} \\
r_{i}^{\,\,j} & \text{\fbox{$x_{ij}$}}%
\end{array}%
\right\vert =x_{ij}-r_{i}^{\,\,j}\big( X^{ij}\big)
^{-1}c_{j}^{\,\,i},
\end{gather*}
where for $1\leq i$, $j\leq N$, $r_{i}^{\,\,j}$ is the row matrix
obtained by removing $j$th entry of $X$ from the $i$th row.
Similarly, $c_{j}^{\,\,i}$ is the column matrix containing $j$th
column of $X$ without $i$th entry.
There exist $N^{2}$ quasideterminants denoted by~$|X|_{ij}$ for $%
i,j=1,\ldots ,N.$ For various properties and applications of
quasideterminants in the theory of integrable systems, see e.g.~\cite{gr3,GR,gr1,gr4,krob}.} (for more details see~\cite{Mhassan})
\begin{gather*}
\psi \lbrack K+1]  = \psi  [ K ] -\Theta  [ K ]
\Lambda
_{K}^{-1}\Theta  [ K ] ^{-1}\psi  [ K ]
 = \left\vert
\begin{array}{cccc}
\Theta _{1} & \cdots & \Theta _{K} & \psi \\
\Theta _{1}\Lambda _{1}^{-1} & \cdots & \Theta _{K}\Lambda _{K}^{-1}
&
\lambda ^{-1}\psi \\
\vdots & \ddots & \vdots & \vdots \\
\Theta _{1}\Lambda _{1}^{-K} & \cdots & \Theta _{K}\Lambda _{K}^{-K}
&
\text{\fbox{$\lambda ^{-K}\psi $}}
\end{array}%
\right\vert .
\end{gather*}
The expression for $S[K+1]$ is given as
\begin{gather*}
S [ K+1 ]  = S-\sum_{l=1}^{K}\Theta  [ K ] \Lambda
_{K}^{-1}\Theta  [ K ] ^{-1}
 = S+\left\vert
\begin{array}{cccc}
\Theta _{1} & \cdots & \Theta _{K} & O \\
\vdots & \ddots & \vdots & \vdots \\
\Theta _{1}\Lambda _{1}^{-\left( K-2\right) } & \cdots & \Theta
_{K}\Lambda
_{K}^{-\left( K-2\right) } & O \\
\Theta _{1}\Lambda _{1}^{-\left( K-1\right) } & \cdots & \Theta
_{K}\Lambda
_{K}^{-\left( K-1\right) } & I \\
\Theta _{1}\Lambda _{1}^{-K} & \cdots & \Theta _{K}\Lambda _{K}^{-K}
&
\text{\fbox{$O$}}
\end{array}
\right\vert .
\end{gather*}
The $K$-fold Darboux transformation on the matrix solution $\psi $
can also be expressed in terms of Hermitian projectors $P[K]$, i.e.
\begin{gather*}
\psi \lbrack K+1]=\prod_{k=0}^{K}\left( I-\frac{\mu _{K-k+1}-{\bar{\mu}%
_{K-k+1}}}{\lambda ^{-1}-{\bar{\mu}_{K-k+1}}}P[K-k+1]\right) \psi ,
\end{gather*}
where the Hermitian projection in this case is
\begin{gather}
P[ k] =\sum_{i=1}^{n}\frac{\vert \theta _{i}[k]
\rangle \langle \theta _{i} [ k ]  \vert }{\langle \theta _{i}[ k] \vert  \theta
_{i}[ k] \rangle },\qquad k=1,2,\dots
,K, \label{projector1}
\end{gather}
with $P^{\dagger }[K]=P[K]$ and $P^{2}[K]=P[K]$.

Now we def\/ine the adjoint Darboux transformation. The equation of motion (\ref{EQOM}) and zero curvature condition (\ref{ZCD}) can also be
written as compatibility condition of the following linear system (the adjoint Lax pair)
\begin{gather}
  {\partial }_{x}\phi  = -{\xi }\partial _{x}S^{\dagger }\phi ,  \qquad
\partial _{t}\phi  = -\big[ S^{\dagger },G^{\dagger }\big] \phi -{\xi } ^{-1}G^{\dagger }\phi ,   \label{LAXAD}
\end{gather}
which is obtained by taking the formal adjoint of the system~(\ref{LAXD}). Note that in equation (\ref{LAXAD}) ${\xi }$ is a real
(or complex)
parameter and $\phi $ is an invertible $n\times n$ matrix in the space $%
\mathcal{V}^{\dag }= \{ \phi  \} $. The Darboux matrix
$D( \xi) $ transforms the matrix solution $\phi $ in space $\mathcal{\tilde{V}}^{\dagger }$ to a new matrix solution $\tilde{\phi}$ in $\mathcal{\tilde{V}}^{\dagger }$, i.e.
\begin{gather}
D ( \xi  ) : \ \mathcal{V}^{\dagger }\rightarrow \mathcal{\tilde{V}}^{\dagger }.  \label{xitrans}
\end{gather}
The one-fold Darboux transformation on the matrix solution $\phi $
is
def\/ined as
\begin{gather*}
\tilde{\phi}\equiv D ( \xi  ) \phi =- \big( \xi
^{-1}I-\Omega \Xi \Omega ^{-1}\big) \phi ,  
\end{gather*}%
where $\Xi =\operatorname{diag}( \xi _{1},\dots ,\xi _{n} ) $ is the
eigenvalue matrix. The matrix function $\Omega $ is an invertible
non-degenerate $n\times n$ matrix and is given by%
\begin{gather*}
\Omega =( \phi ({\xi }_{1})\vert 1\rangle ,\dots ,\phi ({\xi }_{n})\vert n\rangle ) =( \vert \rho
_{1}\rangle ,\dots ,\vert \rho _{n}\rangle ).
\end{gather*}
The $K$-fold Darboux transformation on matrix solutions $\phi$, $S^{\dagger }$ and $G^{\dagger }$ can be expressed as
\begin{gather*}
\phi [ K+1]  = \left\vert
\begin{array}{cccc}
\Omega _{1} & \cdots & \Omega _{K} & \phi \\
\Omega _{1}\Xi _{1}^{-1} & \cdots & \Omega _{K}\Xi _{K}^{-1} & \xi
^{-1}\phi
\\
\vdots & \ddots & \vdots & \vdots \\
\Omega _{1}\Xi _{1}^{-K} & \cdots & \Omega _{K}\Xi _{K}^{-K} & \text{\fbox{$%
\xi ^{-K}\phi $}}
\end{array}
\right\vert , \nonumber\\ 
S^{\dagger }\left[ K+1\right]  = S^{\dagger }+\left\vert
\begin{array}{cccc}
\Omega _{1} & \cdots & \Omega _{K} & O \\
\vdots & \ddots & \vdots & \vdots \\
\Omega _{1}\Xi _{1}^{-\left( K-2\right) } & \cdots & \Omega _{K}\Xi
_{K}^{-\left( K-2\right) } & O \\
\Omega _{1}\Xi _{1}^{-\left( K-1\right) } & \cdots & \Omega _{K}\Xi
_{K}^{-\left( K-1\right) } & I \\
\Omega _{1}\Xi _{1}^{-K} & \cdots & \Omega _{K}\Xi _{K}^{-K} & \text{\fbox{$O$}}%
\end{array}%
\right\vert ,  \qquad 
G^{\dagger }[K+1]  = G.  \nonumber 
\end{gather*}
In terms of the Hermitian projector we write the above expression as%
\begin{gather*}
\phi [K+1]=\prod_{k=0}^{K}\left( I-\frac{\nu _{K-k+1}-\bar{\nu}{_{K-k+1}}}{\xi ^{-1}-\bar{\nu}{_{K-k+1}}}P[K-k+1]\right) \phi ,
\end{gather*}%
and the Hermitian projector in this case is def\/ined as%
\begin{gather}
P [ k ] =\sum_{i=1}^{n}\frac{\vert \rho _{i} [
k ]
 \rangle  \langle \rho _{i} [ k ]  \vert }{%
\langle \rho _{i} [ k ]  \vert   \rho _{i} [ k ]  \rangle },\qquad k=1,2,\dots ,K.
\label{projector2}
\end{gather}
By making use of equations (\ref{LAXD}) and (\ref{LAXAD}) for the
column
solutions~$\vert \theta _{i}\rangle $ and the row solutions~$\langle \rho _{i}\vert $ of the direct and adjoint Lax
pair respectively, it can be easily shown that the expressions~(\ref{projector1}) and~(\ref{projector2}) are equivalent.

\section{Standard binary Darboux transformation}

To def\/ine the binary transformation we follow the approach of \cite{DT,Nimmo1, Nimmo,NGO,OS,rogers} and consider a~space~$\mathcal{\hat{V}}$, which is a
copy of
the direct space $\mathcal{V}$ and the corresponding solutions are~$\hat{\psi}\in \mathcal{\hat{V}}$. Since it is a copy of the direct space,
therefore the linear system, equation of motion and the zero
curvature condition will
have the similar form as given for the direct space. The equation of motion~(\ref{EQOM}) and zero curvature condition~(\ref{ZCD}) can also be
written as the compatibility condition of the following linear
system for the matrix solution~$\hat{\psi}$
\begin{gather}
\ {\partial }_{x}\hat{\psi}  = \hat{U} ( x,t,\lambda  )
\hat{\psi},
\qquad
\partial _{t}\hat{\psi}  = \hat{V} ( x,t,\lambda  ) \hat{\psi},
\label{hatsys}
\end{gather}
where
\begin{gather*}
\hat{U} ( x,t,\lambda  )   = \lambda \partial _{x}\hat{S}, \qquad
\hat{V} ( x,t,\lambda  )   = \big[ \hat{S},\hat{G} \big]
+\lambda ^{-1}\hat{G}.
\end{gather*}
We have taken the specif\/ic solutions $\Theta$, $\Omega $ for the
direct and adjoint spaces $\mathcal{V}$ and $\mathcal{V}^{\dagger }$
respectively. The corresponding solutions for $\mathcal{\hat{V}}$
are $\hat{\Theta}\in \mathcal{\hat{V}}$ and $\hat{\phi}\in
\mathcal{\hat{V}}^{\dagger }$. Also assuming that $i(\hat{\Theta}) \in \mathcal{\tilde{V}}^{\dagger } $, then from
equations (\ref{directtrans}) and (\ref{xitrans}), we write the
transformation as
\begin{gather*}
D^{( -1) \dagger }( \lambda )
: \ \mathcal{V}^{\dagger }\longrightarrow \mathcal{\tilde{V}}^{\dagger}.
\end{gather*}
Since $\phi \in \mathcal{V}^{\dagger }$, we have
\begin{gather*}
i ( \hat{\Theta} ) =D^{( -1) \dagger }(\lambda ) \phi .  
\end{gather*}
Also from $D^{\dagger }( \lambda ) ( i( \Theta
) ) =0$, we obtain $i( \Theta ) =\Theta
^{^{( -1)
\dagger }}$ and similarly $i( \hat{\Theta}) =\hat{\Theta}^{^{( -1) \dagger }}$. Therefore we get from above equation
\begin{gather*}
\hat{\Theta}^{^{ ( -1 ) \dagger }}  = D^{( -1)
\dagger
}( \lambda ) \phi ,  \qquad
\hat{\Theta}  = \big( D^{( -1) \dagger }( \lambda
) \phi \big) ^{( -1) \dagger }. 
\end{gather*}
By using (\ref{darbouxm1}) and (\ref{SM1}) in above equation
\begin{gather}
\hat{\Theta}  = \big( \big( \lambda ^{-1}I-\Theta \Lambda
^{-1}\Theta ^{-1}\big) ^{( -1) \dagger }\phi \big)
^{( -1)\dagger }
 = \big( \lambda ^{-1}I-\Theta \Lambda ^{-1}\Theta ^{-1}\big)
\phi
^{ ( -1)   }  \notag \\
\hphantom{\hat{\Theta}}{}
 = \Theta \big( \lambda ^{-1}I-\Lambda ^{-1}\big) \Theta
^{-1}\phi
^{ ( -1 ) \dagger }
 = \Theta \big( \lambda ^{-1}I-\Lambda ^{-1}\big) \big( \phi
^{\dagger
}\Theta \big) ^{-1}
 = \Theta \Delta ^{-1},  \label{thetaomega}
\end{gather}
where the potential $\Delta $ is def\/ined as
\begin{gather}
\Delta ( \psi ,\phi ) =\big( \phi ^{\dagger }\Theta
\big) \big( \lambda ^{-1}I-\Lambda ^{-1}\big) ^{-1}.
\label{omega}
\end{gather}%
Similarly for adjoint space
\begin{gather*}
\overset{\wedge }{\Omega }=\Omega \Delta ^{ ( -1 ) \dagger
},
\end{gather*}
we obtain%
\begin{gather}
\Delta  ( \psi ,\Omega  ) =-\big( \lambda ^{-1}I-\Xi
^{ ( -1 ) \dagger }\big) ^{-1}\big( \Omega ^{\dagger
}\psi \big) . \label{adomega}
\end{gather}%
By writing equations (\ref{omega}) and (\ref{adomega}) in matrix
form for
the solutions $\Theta $ and $\Omega $, we get the following condition on $\Delta $
\begin{gather}
\Xi ^{( -1) \dagger }\Delta ( \Theta ,\Omega )
-\Delta ( \Theta ,\Omega ) \Lambda ^{-1}=\Omega ^{\dagger
}\Theta , \label{condition}
\end{gather}%
where $\Delta $ is a matrix. An entry $\Delta _{ij}$ from equations (\ref%
{omega}), (\ref{adomega}) and (\ref{condition}) is given as%
\begin{gather}
\Delta ( \Theta ,\Omega ) _{ij}=\frac{( \Omega
^{\dagger }\Theta ) _{ij}}{\bar{\xi}_{i}^{-1}-\lambda
_{j}^{-1}}.  \label{ent}
\end{gather}%
Now we def\/ine the Darboux matrix in hat space as%
\begin{gather}
\hat{D}( \lambda ) \equiv \big( \lambda
^{-1}I-\hat{S}\big)
=\big( \lambda ^{-1}I-\hat{\Theta}\Xi ^{ ( -1 ) \dagger }\hat{\Theta}^{-1}\big) ,  \label{dhat}
\end{gather}%
where
\begin{gather*}
\hat{D} ( \lambda  ) \hat{\psi}=\tilde{\psi}.
\end{gather*}
We may summarize the above formulation as
\begin{gather}
D ( \lambda  )   : \ \mathcal{V}\longrightarrow
\mathcal{\tilde{V}},\nonumber
\\
\hat{D} ( \lambda  )   : \ \mathcal{\hat{V}}\longrightarrow \mathcal{\tilde{V}},  \label{TI} \\
D ( \xi  )   :  \ \mathcal{V}^{\dagger }\longrightarrow \mathcal{\tilde{V}}^{\dagger }.  \nonumber
\end{gather}%
The ef\/fect of $\hat{D}( \lambda ) $ is such that it
leaves the
linear system (\ref{hatsys}) invariant, i.e.,
\begin{gather*}
  {\partial }_{x}\widetilde{\hat{\psi}}  = \widetilde{\hat{U}} (
x,t,\lambda  ) \widetilde{\hat{\psi}},  \qquad
\partial _{t}\widetilde{\hat{\psi}}  = \widetilde{\hat{V}} ( x,t,\lambda
 ) \widetilde{\hat{\psi}},
\end{gather*}
where $\widetilde{\hat{U}}$ and $\widetilde{\hat{V}}$ are given as
\begin{gather*}
\widetilde{\hat{U}} ( x,t,\lambda  )   = \lambda \partial _{x}
\widetilde{\hat{S}}, \qquad
\widetilde{\hat{V}} ( x,t,\lambda  )   = \big[ \widetilde{\hat{S}},
\widetilde{\hat{G}}\big] +\lambda ^{-1}\widetilde{\hat{G}},
\end{gather*}
and $\tilde{S}$ and $\tilde{G}$ are the Lie algebra valued
transformed
matrix f\/ields. By substituting equation~(\ref{Vd}) in equations~(\ref{LAXD1}), we get
\begin{gather*}
\widetilde{\hat{S}}  = \hat{S}-\hat{M}, \qquad
\widetilde{\hat{G}}  = \hat{G}.
\end{gather*}
As mentioned earlier equation (\ref{G1}) shows that $\hat{G}$ is a
constant matrix and the matrix $M$ is subjected to satisfy the
following equations
\begin{gather*}
\partial _{x}\hat{M}\hat{M}  =\big[ \partial _{x}\hat{S},\hat{M}\big] ,
\qquad
\partial _{t}\hat{M}  =\big[ \big[ \hat{S},\hat{G}\big] ,\hat{M}\big]
+\big[ \hat{G},\hat{M}\big] \hat{M}.
\end{gather*}
The matrix $M$ can be written in terms of the solutions of the
linear system
\begin{gather*}
\hat{M}=\hat{\Theta}\Xi ^{ ( -1 ) \dagger
}\hat{\Theta}^{-1},
\end{gather*}
and the Darboux transformation on the matrix f\/ields $\hat{\psi}$ and~$\hat{S}
$ in hat space~$\mathcal{\hat{V}}$ is
\begin{gather*}
\widetilde{\hat{\psi}}  = \big(\lambda ^{-1}I-\hat{\Theta}\Xi ^{ (
-1 )
\dagger }\hat{\Theta}^{-1}\big)\hat{\psi},  \qquad 
\widetilde{\hat{S}}  = \hat{S}-\hat{\Theta}\Xi ^{ ( -1 ) \dagger } \hat{\Theta}^{-1}.  
\end{gather*}
From equation (\ref{TI}) we know that%
\begin{gather*}
\hat{D} ( \lambda  ) \hat{\psi}=D ( \lambda  )
\psi ,
\end{gather*}
which implies
\begin{gather}
\hat{\psi}=\hat{D}^{-1} ( \lambda  ) D ( \lambda
 ) \psi . \label{Uhat}
\end{gather}%
The equation (\ref{Uhat}) relates the two solutions $\psi $ and
$\hat{\psi}$. This transformation is known as the standard binary
Darboux transformation and we write it as $B ( \lambda  )
=\hat{D}^{-1} ( \lambda
 ) D ( \lambda  ) $, i.e.
\begin{gather}
\hat{\psi}=\hat{D}^{-1} ( \lambda  ) D ( \lambda
 ) \psi =B ( \lambda  ) \psi .  \label{binary}
\end{gather}%
By substituting (\ref{dhat}), (\ref{darbouxm1}) in equation
(\ref{binary}), we obtain the explicit transformation on $\psi $ as
\begin{gather}
\hat{\psi}  = \big( \lambda ^{-1}I-\hat{\Theta}\Xi ^{ (
-1 ) \dagger }\hat{\Theta}^{-1}\big) ^{-1}\big( \lambda
^{-1}I-\Theta \Lambda
^{-1}\Theta ^{-1}\big) \psi    \notag \\
\hphantom{\hat{\psi}}{} = \hat{\Theta}\big( \lambda ^{-1}I-\Xi ^{ ( -1 ) \dagger
}\big) ^{-1}\hat{\Theta}^{-1}\Theta \big( \lambda ^{-1}I-\Lambda
^{-1}\big) \Theta ^{-1}\psi .  \label{binary2}
\end{gather}
By using (\ref{thetaomega}) in equation (\ref{binary2}), the expression of $%
\hat{\psi}$ may be simplif\/ied as
\begin{gather*}
\hat{\psi}  = \Theta \Delta  ( \Theta ,\Omega  )
^{-1}\big( \lambda ^{-1}I-\Xi ^{ ( -1 ) \dagger }\big)
^{-1}\Delta  ( \Theta ,\Omega  ) \Theta ^{-1}\Theta \big(
\lambda ^{-1}I-\Lambda
^{-1}\big) \Theta ^{-1}\psi  \\
\hphantom{\hat{\psi}}{}
 = \Theta \Delta  ( \Theta ,\Omega  ) ^{-1}\big( \lambda
^{-1}I-\Xi ^{ ( -1 ) \dagger }\big) ^{-1}\Delta  (
\Theta ,\Omega  ) \big( \lambda ^{-1}I-\Lambda ^{-1}\big)
\Theta ^{-1}\psi
\\
\hphantom{\hat{\psi}}{}
 = \Theta \Delta  ( \Theta ,\Omega  ) ^{-1}\big( \lambda
^{-1}I-\Xi ^{ ( -1 ) \dagger }\big) ^{-1}\big( \lambda
^{-1}\Delta  ( \Theta ,\Omega  ) -\Delta  ( \Theta
,\Omega  ) \Lambda ^{-1}\big) \Theta ^{-1}\psi .
\end{gather*}%
By substituting the value of $\Delta  ( \Theta ,\Omega  )
\Lambda
^{-1}$ from (\ref{condition}), we get
\begin{gather}
\hat{\psi}  = \Theta \Delta  ( \Theta ,\Omega  )
^{-1}\big( \lambda ^{-1}I-\Xi ^{ ( -1 ) \dagger }\big)
^{-1}\big( \lambda ^{-1}\Delta  ( \Theta ,\Omega  ) -\Xi
^{ ( -1 ) \dagger }\Delta  ( \Theta ,\Omega  )
+\Omega ^{\dagger }\Theta \big)
\Theta ^{-1}\psi   \notag \\
\hphantom{\hat{\psi}}{}
 = \big( I+\Theta \Delta  ( \Theta ,\Omega  ) ^{-1}\big(
\lambda ^{-1}I-\Xi ^{ ( -1 ) \dagger }\big) ^{-1}\Omega
^{\dagger
}\big) \psi
 = \big( I-\Theta \Delta  ( \Theta ,\Omega  ) ^{-1}\Delta
 (
\cdot ,\Omega  ) \big) \psi   \notag \\
\hphantom{\hat{\psi}}{}
 = \psi -\Theta \Delta  ( \Theta ,\Omega  ) ^{-1}\Delta
 ( \psi ,\Omega  ) ,  \label{Binary1}
\end{gather}
where we have used equation (\ref{adomega}) in obtaining the last
step.
Equation (\ref{Binary1}) may be written in terms of quasideterminant as
\begin{gather}
\hat{\psi}=\left\vert
\begin{array}{cc}
\Delta \left( \Theta ,\Omega \right)  & \Delta \left( \psi ,\Omega
\right)
\\
\Theta  & \text{\fbox{$\psi $}}%
\end{array}%
\right\vert .  \label{psi}
\end{gather}%
The quasideterminant (\ref{psi}) is referred to as quasi-Grammian
solution
of the system. The adjoint binary transformation for $\hat{\phi}\in \mathcal{%
\hat{V}}^{\dagger }$ is obtained in a simmilar way and gives%
\begin{gather*}
\hat{\phi}  = \phi -\Omega \Delta  ( \Theta ,\Omega  )
^{ (
-1 ) \dagger }\Delta ^{\dagger } ( \Theta ,\phi  )
 = \left\vert
\begin{array}{cc}
\Delta ^{\dagger } ( \Theta ,\Omega  )  & \Delta ^{\dagger
} (
\Theta ,\phi  )  \\
\Omega  & \text{\fbox{$\phi $}}%
\end{array}%
\right\vert .
\end{gather*}%
Again from equation (\ref{Uhat}), we have%
\begin{gather*}
\hat{S}-\hat{\Theta}\Xi ^{ ( -1 ) \dagger
}\hat{\Theta}^{-1}
 = S-\Theta \Lambda ^{-1}\Theta ^{-1}, \\
\hat{S}  = S-\Theta \Lambda ^{-1}\Theta ^{-1}+\hat{\Theta}\Xi
^{ (
-1 ) \dagger }\hat{\Theta}^{-1}
 = S-\Theta \Lambda ^{-1}\Theta ^{-1}+\Theta \Delta  ( \Theta
,\Omega  ) ^{-1}\Xi ^{ ( -1 ) \dagger }\Delta  (
\Theta ,\Omega  ) \Theta ^{-1}.
\end{gather*}
By using equation (\ref{condition}) for $\Xi ^{\left( -1\right)
\dagger }\Delta \left( \Theta ,\Omega \right) $ in above equation
\begin{gather*}
\hat{S}  = S-\Theta \Lambda ^{-1}\Theta ^{-1}+\Theta \Delta  (
\Theta ,\Omega  ) ^{-1}\big( \Delta  ( \Theta ,\Omega
 ) \Lambda
^{-1}+\Omega ^{\dagger }\Theta \big) \Theta ^{-1} \\
\hphantom{\hat{S}}{}
 = S+\Theta \Delta  ( \Theta ,\Omega  ) ^{-1}\Omega ^{\dagger }
 = S-\left\vert
\begin{array}{cc}
\Delta  ( \Theta ,\Omega  )  & \Omega ^{\dagger } \\
\Theta  & \text{\fbox{$O$}}%
\end{array}%
\right\vert .
\end{gather*}
For the next iteration of binary Darboux transformation, we take
$\Theta _{1} $, $\Theta _{2}$ to be two particular solutions of the
Lax pair (\ref{LAXD}) at $\Lambda =\Lambda _{1}$ and $\Lambda
=\Lambda _{2}$ respectively. Similarly $\Omega _{1}$, $\Omega _{2}$ are
two particular solutions of the Lax pair (\ref{LAXAD}) at $\Xi =\Xi
_{1}$ and $\Xi =\Xi _{2}$. Using the
notation $\psi  [ 1 ] =\psi$, $S [ 1 ] =S$ and $\psi  [ 2
 ] =\hat{\psi}$, $S [ 2 ] =\hat{S}$, we write two-fold
binary
Darboux transformation on $\psi $ as
\begin{gather}
\psi  [ 3 ] =\psi  [ 2 ] -\Theta  [ 2 ]
\Delta  ( \Theta  [ 2 ] ,\Omega [ 2 ]  )
^{-1}\Delta  ( \psi  [ 2 ] ,\Omega  [ 2 ]
 ) ,  \label{psi3}
\end{gather}%
where $\Theta  [ 1 ] =\Theta _{1}$, $\Omega  [ 1 ]
=\Omega _{1}$, $\Theta  [ 2 ] = \psi  [ 2 ]
 \vert _{\psi \rightarrow \Theta _{2}}$, $\Omega  [ 2 ]
=  \phi  [ 2 ]  \vert _{\phi \rightarrow \Omega
_{2}}$. Also note that by using the
def\/inition of the potential $\Delta $ and equation~(\ref{ent}), we have
\begin{gather}
\Delta  ( \psi  [ 2 ] ,\phi  [ 2 ]  )
 = \Delta  ( \psi _{1},\phi _{1} ) -\Delta  ( \Theta
_{1},\phi _{1} ) \Delta  ( \Theta _{1},\Omega _{1} )
^{-1}\Delta  (
\psi _{1},\Omega _{1} )    \notag \\
\hphantom{\Delta  ( \psi  [ 2 ] ,\phi  [ 2 ]  )}{}
 = \left\vert
\begin{array}{cc}
\Delta  ( \Theta _{1},\Omega _{1} )  & \Delta  ( \psi
,\Omega
_{1} )  \\
\Delta  ( \Theta _{1},\phi  )  & \text{\fbox{$\Delta
 ( \psi
,\phi  ) $}}%
\end{array}%
\right\vert .  \label{omega3}
\end{gather}
The equation (\ref{omega3}) implies that
\begin{gather}
\Delta  ( \Theta  [ 2 ] ,\Omega  [ 2 ]  )
 = \Delta  ( \Theta _{2},\Omega _{2} ) -\Delta  (
\Theta _{1},\Omega _{2} ) \Delta  ( \Theta _{1},\Omega
_{1} )
^{-1}\Delta  ( \Theta _{2},\Omega _{1} )    \notag \\
\hphantom{\Delta  ( \Theta  [ 2 ] ,\Omega  [ 2 ]  )}{}
 = \left\vert
\begin{array}{cc}
\Delta  ( \Theta _{1},\Omega _{1} )  & \Delta  ( \Theta
_{2},\Omega _{1} )  \\
\Delta  ( \Theta _{1},\Omega _{2} )  & \text{\fbox{$\Delta
(
\Theta _{2},\Omega _{2} ) $}}%
\end{array}
\right\vert .  \label{omega2}
\end{gather}
By using equations (\ref{omega3}), (\ref{omega2}) and the notation
def\/ined
above in equation (\ref{psi3}), we get
\begin{gather}
\psi  [ 3 ]   = \left\vert
\begin{array}{cc}
\Delta  ( \Theta _{1},\Omega _{1} )  & \Delta  ( \psi
,\Omega
_{1} )  \\
\Theta _{1} & \text{\fbox{$\psi $}}%
\end{array}%
\right\vert -\left\vert
\begin{array}{cc}
\Delta  ( \Theta _{1},\Omega _{1} )  & \Delta  ( \Theta
_{2},\Omega _{1} )  \\
\Theta _{1} & \text{\fbox{$\Theta _{2}$}}%
\end{array}%
\right\vert   \notag \\
\hphantom{\psi  [ 3 ]   =}{}
\times \left\vert
\begin{array}{cc}
\Delta  ( \Theta _{1},\Omega _{1} )  & \Delta  ( \Theta
_{2},\Omega _{1} )  \\
\Delta  ( \Theta _{1},\Omega _{2} )  & \text{\fbox{$\Delta
 (
\Theta _{2},\Omega _{2} ) $}}%
\end{array}
\right\vert ^{-1}\left\vert
\begin{array}{cc}
\Delta  ( \Theta _{1},\Omega _{1} )  & \Delta  ( \psi
,\Omega
_{1} )  \\
\Delta  ( \Theta _{1},\phi  )  & \text{\fbox{$\Delta
 ( \psi
,\Omega _{2} ) $}}%
\end{array}%
\right\vert   \notag \\
\hphantom{\psi  [ 3 ] }{}
=\left\vert
\begin{array}{ccc}
\Delta  ( \Theta _{1},\Omega _{1} )  & \Delta  ( \Theta
_{2},\Omega _{1} )  & \Delta  ( \psi ,\Omega _{1} )  \\
\Delta  ( \Theta _{1},\Omega _{2} )  & \Delta  ( \Theta
_{2},\Omega _{2} )  & \Delta  ( \psi ,\Omega _{2} )  \\
\Theta _{1} & \Theta _{2} & \text{\fbox{$\psi $}}%
\end{array}%
\right\vert ,  \label{psi3q}
\end{gather}
where we have used the noncommutative Jacobi identity\footnote{For quasideterminants, the noncommutative Jacobi identity is given as
\begin{gather*}
\left\vert
\begin{array}{ccc}
E & F & G \\
H & A & B \\
J & C & \text{\fbox{$D$}}%
\end{array}%
\right\vert =\left\vert
\begin{array}{cc}
E & G \\
J & \text{\fbox{$D$}}%
\end{array}%
\right\vert -\left\vert
\begin{array}{cc}
E & F \\
J & \text{\fbox{$C$}}%
\end{array}%
\right\vert \left\vert
\begin{array}{cc}
E & F \\
H & \text{\fbox{$A$}}%
\end{array}%
\right\vert ^{-1}\left\vert
\begin{array}{cc}
E & G \\
H & \text{\fbox{$B$}}%
\end{array}%
\right\vert .
\end{gather*}
For the def\/inition and more properties of quasideterminants see e.g.~\cite{gr3,GR,gr1,gr4,krob}.} in obtaining~(\ref{psi3q}). The $K$th iteration
of binary
Darboux transformation leads to
\begin{gather}
\psi  [ K+1 ]   = \psi  [ K ] -\Theta  [
K ] \Delta  ( \Theta  [ K ] ,\Omega  [
K ]  ) ^{-1}\Delta  ( \psi  [ K ] ,\Omega
 [ K ]  )
\notag \\
\hphantom{\psi  [ K+1 ]}{}
= \left\vert
\begin{array}{cc}
\Delta  ( \Theta  [ K ] ,\Omega  [ K ]  )
&
\Delta  ( \psi  [ K ] ,\Omega  [ K ]  )  \\
\Theta  [ K ]  & \text{\fbox{$\psi  [ K ] $}}%
\end{array}%
\right\vert    \notag \\
\hphantom{\psi  [ K+1 ]}{}
 = \left\vert
\begin{array}{cccc}
\Delta  ( \Theta _{1},\Omega _{1} )  & \cdots  & \Delta
 (
\Theta _{K},\Omega _{1} )  & \Delta  ( \psi ,\Omega _{1} )  \\
\vdots  & \cdots  & \vdots  & \vdots  \\
\Delta  ( \Theta _{1},\Omega _{K} )  & \cdots  & \Delta
 (
\Theta _{K},\Omega _{K} )  & \Delta  ( \psi ,\Omega _{K} )  \\
\Theta _{1} & \cdots  & \Theta _{K} & \text{\fbox{$\psi $}}%
\end{array}
\right\vert .  \label{psik}
\end{gather}
Above result can be proved by induction by using the properties of
quasideterminants. Similarly the $K$th iteration of adjoint binary
Darboux
transformation gives
\begin{gather*}
\phi  [ K+1 ]   = \phi  [ K ] -\Omega  [
K ] \Delta  ( \Theta  [ K ] ,\Omega  [
K ]  ) ^{( -1 ) \dagger }\Delta  ( \Theta
 [ K ] ,\phi  [
K ]  ) ^{\dagger }   \notag \\
\hphantom{\phi  [ K+1 ]}{}
 = \left\vert
\begin{array}{cc}
\Delta  ( \Theta  [ K ] ,\Omega  [ K ]  )
^{\dagger } & \Delta  ( \Theta  [ K ] ,\phi  [
K ]
 ) ^{\dagger } \\
\Omega  [ K ]  & \text{\fbox{$\phi  [ K ] $}}%
\end{array}%
\right\vert    \notag \\
\hphantom{\phi  [ K+1 ]}{}
 = \left\vert
\begin{array}{ccccc}
\Delta  ( \Theta _{1},\Omega _{1} ) ^{\dagger } & \Delta
 ( \Theta _{2},\Omega _{1} ) ^{\dagger } & \cdots  & \Delta
 ( \Theta _{K},\Omega _{1} ) ^{\dagger } & \Delta  (
\Theta _{1},\phi  )
^{\dagger } \\
\Delta  ( \Theta _{1},\Omega _{2} ) ^{\dagger } & \Delta
 ( \Theta _{2},\Omega _{2} ) ^{\dagger } & \cdots  & \Delta
 ( \Theta _{K},\Omega _{2} ) ^{\dagger } & \Delta  (
\Theta _{2},\phi  )
^{\dagger } \\
\vdots  & \vdots  & \cdots  & \vdots  & \vdots  \\
\Delta  ( \Theta _{1},\Omega _{K} ) ^{\dagger } & \Delta
 ( \Theta _{2},\Omega _{K} ) ^{\dagger } & \cdots  & \Delta
 ( \Theta _{K},\Omega _{K} ) ^{\dagger } & \Delta  (
\Theta _{K},\phi  )
^{\dagger } \\
\Omega _{1} & \Omega _{2} & \cdots  & \Omega _{K} & \text{\fbox{$\phi $}}%
\end{array}%
\right\vert .  
\end{gather*}
The multisoliton $S [ K+1 ] $ can be obtained by putting
$\lambda =0
$ in the expression for $\psi  [ K+1 ] $ (\ref{psik}) and using $G= \psi  \vert _{\lambda =0}$, which on silmplif\/ication gives{\samepage
\begin{gather*}
S [ K+1 ] =S-\left\vert
\begin{array}{cccc}
\Delta  ( \Theta _{1},\Omega _{1} )  & \cdots  & \Delta
 (
\Theta _{K},\Omega _{1} )  & \Omega _{1}^{\dagger } \\
\vdots  & \cdots  & \vdots  & \vdots  \\
\Delta  ( \Theta _{1},\Omega _{K} )  & \cdots  & \Delta
 (
\Theta _{K},\Omega _{K} )  & \Omega _{K}^{\dagger } \\
\Theta _{1} & \cdots  & \Theta _{K} & \text{\fbox{$I$}}%
\end{array}%
\right\vert .
\end{gather*}%
Similar expression can be obtained for the $K$th iteration of~$S^{\dagger }$.}

Therefore by using the standard binary Darboux transformation we
have obtained the grammian type solutions for the linear system and
the potential is also expressed in terms of quasideterminants. That
is by constructing binary Darboux transformation in terms of
spectral para\-me\-ter we can get the expression of the matrix solution
of the linear system in terms of grammian type quasideterminants
which is dif\/ferent in representation from the solutions obtained by
elementary Darboux transformation. In addition to the solutions of
the linear system we are also able to obtain explicit
quasideterminant expression of the potential~$\Delta $ in terms of
the particular solutions of the linear system. It is important to
note that the spectral parameter remains unchanged in binary Darboux
transformation. We consider the eigenfunctions (solutions of direct
Lax pair) and adjoint eigenfunctions (solutions of adjoint pair).
The bilinear potential~$\Delta$ is related to each pair of (direct
and adjoint) solutions. Since we know that the solutions of the
linear system can be column vectors or they can be combined to give
solution in matrix form. As in the present case when the solutions
are in matrix form the potential~$\Delta$ is also a matrix. It has
been shown earlier that matrix solutions can be reduced to vector
solutions~\cite{BDTPCM}. In such a~case when solutions are vectors
the potential~$\Delta$ becomes scalar and by replacing spectral
para\-me\-ter with derivative we can consider potential to be a contour
integration of the corresponding expressions in~$x-t$ plane. In the
next section we will see what happens when we apply our method to a
specif\/ic case of $SU( 2) $ system.

\section[Explicit solutions for the $SU(2) $ system]{Explicit solutions for the $\boldsymbol{SU(2)}$ system}

In this section we consider the generalized coupled dispersionless
integrable system based on the Lie group $SU(2)$ and calculate the
soliton
solutions by using binary Darboux transformation. For the Lie group $SU( 2) $ the matrix f\/ields $S$ and $G$ are valued in the
Lie
algebra $\mathbf{su}( 2) $ and we have%
\begin{alignat}{3}
& S^{\dagger }  = -S,\qquad && G^{\dagger
}=-G,&
\label{Herm} \\
&\operatorname{Tr} S  = 0,\qquad &&  \operatorname{Tr} G=0.& \label{Tr}
\end{alignat}
Following the same steps for the direct Lax pair as obtained in \cite{Mhassan}. We def\/ine a vector $\phi = ( \phi _{1},\phi _{2},\phi
_{3} ) $ in such a way that the matrix f\/ield $S$ is given by
\begin{gather}
S=i
\begin{pmatrix}
\phi _{3} & \phi _{1}-i\phi _{2} \\
\phi _{1}+i\phi _{2} & -\phi _{3}
\end{pmatrix}.  \label{SDIS}
\end{gather}%
Equation (\ref{SDIS}) satisf\/ies the conditions (\ref{Herm}) and
(\ref{Tr}).
The matrices $U$ and $V$ are then given as%
\begin{gather*}
U  = i\lambda
\begin{pmatrix}
\partial _{x}\phi _{3} & \partial _{x}\phi _{1}-i\partial _{x}\phi _{2} \\
\partial _{x}\phi _{1}+i\partial _{x}\phi _{2} & -\partial _{x}\phi _{3}
\end{pmatrix} , \qquad\!
V  =
\begin{pmatrix}
0 & \phi _{1}-i\phi _{2} \\
-\phi _{1}-i\phi _{2} & 0%
\end{pmatrix} -\frac{i}{2\lambda } \!
\begin{pmatrix}
1 & 0 \\
0 & -1%
\end{pmatrix} .
\end{gather*}
By writing $\phi _{1}=r$, $\phi _{2}=0$ and $\phi _{3}=q$, we get the
coupled
dispersionless integrable system as given in \cite{konno}
\begin{gather*}
\partial _{x}\partial _{t}q+2\partial _{x}rr  = 0, \qquad
\partial _{x}\partial _{t}r-2\partial _{x}qr  = 0.
\end{gather*}
and the matrix $S$ from equation (\ref{SDIS}) is given as
\begin{gather}
S=i
\begin{pmatrix}
q & r \\
r & -q%
\end{pmatrix} .  \label{SQR}
\end{gather}%
To obtain the expression for the Darboux matrix we take
\begin{gather}
\Lambda  =
\begin{pmatrix}
\lambda _{1} & 0 \\
0 & -\lambda _{1}%
\end{pmatrix} , \qquad  
\Theta  =
\begin{pmatrix}
\alpha & \beta \\
\beta & -\alpha%
\end{pmatrix}.  \label{THETA1}
\end{gather}
By using equations 
(\ref{THETA1}) and (\ref{SM1}) in
equation (\ref{darbouxm1}), we get
\begin{gather*}
D ( \lambda  ) =
\begin{pmatrix}
\lambda ^{-1}-\lambda _{1}^{-1}\cos \omega & -\lambda _{1}^{-1}\sin
\omega
\\
-\lambda _{1}^{-1}\sin \omega & \lambda ^{-1}-\lambda _{1}^{-1}\cos \omega%
\end{pmatrix} ,
\end{gather*}
where we have assumed that $\tan \frac{\omega }{2}=\frac{\beta
}{\alpha }$.
We now consider the seed solution as follows
\begin{gather}
\psi =
\begin{pmatrix}
e^{i\lambda x-\frac{i}{2\lambda }t} & 0 \\
0 & e^{-i\lambda x+\frac{i}{2\lambda }t}%
\end{pmatrix}.  \label{Psi matrix}
\end{gather}%
By using the above equation (\ref{Psi matrix}) we can write the
particular matrix solution $\Theta $ of the direct Lax pair
(\ref{LAXD}) as
\begin{gather}
\Theta  =
\begin{pmatrix}
\psi  ( \lambda _{1} )  \vert 1 \rangle & \psi
 (
\lambda _{2} )  \vert 2 \rangle%
\end{pmatrix}
 =
\begin{pmatrix}
e^{i\lambda _{1}x-\frac{i}{2\lambda _{1}}t} & e^{i\lambda _{2}x-\frac{i}{%
2\lambda _{2}}t} \\
e^{-i\lambda _{1}x+\frac{i}{2\lambda _{1}}t} & -e^{-i\lambda _{2}x+\frac{i}{%
2\lambda _{2}}t}%
\end{pmatrix} .  \label{THETAL}
\end{gather}%
By substituting $\lambda _{2}=-\lambda _{1}$ we get from above equation (\ref{THETAL})
\begin{gather}
\Theta =
\begin{pmatrix}
e^{i\lambda _{1}x-\frac{i}{2\lambda _{1}}t} & e^{-i\lambda _{1}x+\frac{i}{%
2\lambda _{1}}t} \\
e^{-i\lambda _{1}x+\frac{i}{2\lambda _{1}}t} & -e^{i\lambda _{1}-\frac{i}{%
2\lambda _{1}}t}%
\end{pmatrix} .  \label{THETAL1}
\end{gather}%
Taking $l=2i\lambda _{1}x-\frac{i}{\lambda _{1}}t$ and using the def\/inition (\ref{SM1}) we get from (\ref{THETAL1})%
\begin{gather}
M  = \frac{\lambda _{1}^{-1}}{2\cosh l}
\begin{pmatrix}
2\sinh l & 2 \\
2 & -2\sinh l%
\end{pmatrix}
 = \lambda _{1}^{-1}
\begin{pmatrix}
\tanh l & \operatorname{sech}l \\
\operatorname{sech} l & -\tanh l%
\end{pmatrix} .  \label{MR}
\end{gather}
From equation (\ref{S1}) we have%
\begin{gather*}
\partial _{x}S [ 1 ] =\partial _{x}S-\partial _{x}M.
\end{gather*}
On comparison with equation (\ref{SQR}) and using (\ref{MR}) we obtain%
\begin{gather}
\partial _{x}q [ 1 ]  = \partial _{x}q+i\partial _{x}M_{11}
=1+i\lambda _{1}^{-1}\partial _{x}\tanh l
 = 1-2\operatorname{sech}^{2}l \nonumber\\
 \hphantom{\partial _{x}q [ 1 ]}{}
 = 1-2\operatorname{sech}^{2}\left( 2i\lambda _{1}x-\frac{i}{\lambda _{1}}%
t\right) ,  \label{DXQR}
\end{gather}
where we have used $\partial _{x}q=1$. Similarly we have for $r=0$%
\begin{gather*}
\partial _{x}r [ 1 ] =\partial _{x}r+i\partial _{x}M_{12},
\end{gather*}
which gives%
\begin{gather}
r [ 1 ]  = iM_{12}
 = i\lambda _{1}^{-1}\operatorname{sech}\left( 2i\lambda
_{1}x-\frac{i}{\lambda _{1}}t\right) .  \label{RX1}
\end{gather}
It is easy to show from equations (\ref{DXQR}) and (\ref{RX1}) that
in the
asymptotic limit $\partial _{x}q [ 1 ] \rightarrow 1$ and $r [ 1] \rightarrow 0$. Now by making use of above calculation we
can write
the iterated solution $\psi  [ 1 ] $ as%
\begin{gather*}
\psi  [ 1 ] =D ( \lambda  ) \psi =
\begin{pmatrix}
\left( \lambda ^{-1}-\lambda _{1}^{-1}\tanh l\right) e^{\frac{l}{2}}
&
-\lambda _{1}^{-1}\operatorname{sech} le^{\frac{-l}{2}} \\
-\lambda _{1}^{-1}\operatorname{sech}e ^{\frac{l}{2}} & \left( \lambda
^{-1}+\lambda _{1}^{-1}\tanh l\right) e^{\frac{-l}{2}}%
\end{pmatrix} .  
\end{gather*}
Repeating the calculations as we did for direct pair, we get%
\begin{gather}
\Omega =
\begin{pmatrix}
e^{\frac{p}{2}} & e^{-\frac{p}{2}} \\
e^{-\frac{p}{2}} & -e^{\frac{p}{2}}%
\end{pmatrix},  \label{omegamatrix11}
\end{gather}%
where
$
p=2i\xi _{1}x-\frac{i}{\xi _{1}}t$.  
To obtain the expression for $\hat{S}$, we start with the def\/inition (\ref{ent}) of $\Delta  ( \Theta ,\Omega  )$, $\bar{\xi}=-\xi $
and by
using (\ref{THETAL1}), (\ref{omegamatrix11}) obtain for the present case%
\begin{gather*}
\Delta  ( \Theta ,\Omega  ) =
\begin{pmatrix}
-\dfrac{2\cosh \hat{l}}{\xi _{1}^{-1}+\lambda _{1}^{-1}} & -\dfrac{2\sinh \hat{%
p}}{\xi _{1}^{-1}-\lambda _{1}^{-1}} \vspace{1mm}\\
-\dfrac{2\sinh \hat{p}}{\xi _{1}^{-1}-\lambda _{1}^{-1}} & \dfrac{2\cosh \hat{l%
}}{\xi _{1}^{-1}+\lambda _{1}^{-1}}%
\end{pmatrix} ,  
\end{gather*}%
where
\begin{gather*}
\hat{l}(x^{+},x^{-})  = i ( \xi _{1}+\lambda _{1} ) x-\frac{i}{2}
\left( \frac{1}{\xi _{1}}+\frac{1}{\lambda _{1}}\right) t,  \qquad
\hat{p}(x^{+},x^{-})  = i ( \xi _{1}-\lambda _{1} ) x-\frac{i}{2}
\left( \frac{1}{\xi _{1}}-\frac{1}{\lambda _{1}}\right) t.
\end{gather*}%
Now we consider
\begin{gather}
\hat{M}  = \Theta \Delta  ( \Theta ,\Omega  ) ^{-1}\Omega
^{\dagger }=
\begin{pmatrix}
\hat{M}_{11} & \hat{M}_{12} \\
\hat{M}_{21} & \hat{M}_{22}
\end{pmatrix}  \notag \\
\hphantom{\hat{M}}{}
 = \frac{4}{K}
\begin{pmatrix}
\dfrac{\cosh \hat{l}\sinh \hat{l}}{\xi _{1}^{-1}+\lambda _{1}^{-1}}+\dfrac{%
\cosh \hat{p}\sinh \hat{p}}{\xi _{1}^{-1}-\lambda _{1}^{-1}} &
\dfrac{\cosh
\hat{l}\cosh \hat{p}}{\xi _{1}^{-1}+\lambda _{1}^{-1}}-\dfrac{\sinh \hat{p}%
\sinh \hat{l}}{\xi _{1}^{-1}-\lambda _{1}^{-1}} \vspace{1mm}\\
\dfrac{\cosh \hat{l}\cosh \hat{p}}{\xi _{1}^{-1}+\lambda _{1}^{-1}}-\dfrac{%
\sinh \hat{p}\sinh \hat{l}}{\xi _{1}^{-1}-\lambda _{1}^{-1}} &
-\dfrac{\cosh
\hat{l}\sinh \hat{l}}{\xi _{1}^{-1}+\lambda _{1}^{-1}}-\dfrac{\cosh \hat{p}%
\sinh \hat{p}}{\xi _{1}^{-1}-\lambda _{1}^{-1}}%
\end{pmatrix} ,  \label{Mhat}
\end{gather}
where
\begin{gather}
K=\det \Delta  ( \Theta ,\Omega  ) =\frac{-4\cosh ^{2}\hat{l}}{ \left( \xi _{1}^{-1}+\lambda _{1}^{-1}\right) ^{2}}-\frac{4\sinh ^{2}\hat{p}%
}{\left( \xi _{1}^{-1}-\lambda _{1}^{-1}\right) ^{2}},  \nonumber 
\qquad
\hat{S}  = S+\Theta \Delta  ( \Theta ,\Omega  ) ^{-1}\Omega
^{\dagger },  \notag \\
\begin{pmatrix}
iq [ 1 ] & ir [ 1 ] \\
ir [ 1 ] & -iq [ 1 ]%
\end{pmatrix}  =
\begin{pmatrix}
iq+\hat{M}_{11} & ir+\hat{M}_{12} \\
ir+\hat{M}_{21} & -iq+\hat{M}_{22}
\end{pmatrix} \label{Shat}
\end{gather}
From equation (\ref{Shat}) and (\ref{Mhat}) by using $\partial _{x}q=1$ and $r=0$ we get%
\begin{gather}
\partial _{x}q [ 1 ]  = 1+\frac{8\lambda \xi }{K}\left[ \sinh
p\sinh l+\frac{2}{K}\left\{ \frac{\sinh 2\hat{l}}{\xi
_{1}^{-1}+\lambda
_{1}^{-1}}-\frac{\sinh 2\hat{p}}{\xi _{1}^{-1}-\lambda _{1}^{-1}}\right\} %
\right] ,  \label{Q1} \\
r [ 1 ]  = -i\frac{4}{K}\left\{ \frac{\cosh \hat{l}\cosh \hat{p}}{%
\xi _{1}^{-1}+\lambda _{1}^{-1}}-\frac{\sinh \hat{p}\sinh
\hat{l}}{\xi _{1}^{-1}-\lambda _{1}^{-1}}\right\} .  \label{R1}
\end{gather}

In the asymptotic limit for $t\rightarrow \pm \infty $, we have $\hat{l}%
\rightarrow \pm \infty $ and the equations~(\ref{Q1}) and~(\ref{R1}) become
\begin{gather}
\lim_{\hat{l}\rightarrow \pm \infty }\partial _{x}q [ 1 ]   = 1, \qquad
\lim_{\hat{l}\rightarrow \pm \infty }r [ 1 ]   = 0.
\label{sprimeneu}
\end{gather}
We see that in the asymptotic limit, we get much simpler
expressions. Note that the expression is similar to the one we
obtain from elementary Darboux transformation. Now we consider the
special case when $\xi =\lambda $ which
gives $\hat{p}=0$. The solutions~(\ref{Q1}) and~(\ref{R1}) become%
\begin{gather}
\partial _{x}q [ 1 ]   = 1+2\operatorname{sech}^{2}\hat{l},  \label{R1e}
\\
r [ 1 ]   = i\lambda ^{-1}\operatorname{sech}\hat{l}.
\label{Q1e}
\end{gather}
On comparison of equations (\ref{DXQR}) and (\ref{RX1}) with equations~(\ref{Q1}) and~(\ref{R1}) we see that the original solutions obtained by
the standard binary Darboux transformation are dif\/ferent from those
of elementary Darbouix transformation and contain the contribution
from both the direct and adjoint system. If we take $\xi =\lambda $
the solutions from
both the techniques become equal as shown by equations~(\ref{R1e}) and~(\ref{Q1e}). Therefore the advantage of using standard binary Darboux
transformation is that we can obtain the solutions in the form of
direct and adjoint space parameters and then without using
elementary Darboux transformation we can obtain solutions just by
equating parameters as shown above where we have obtained equations
(\ref{R1e}) and~(\ref{Q1e}) (which have same form as solutions
obtained from elementary Darboux transformation) from equations~(\ref{Q1}) and~(\ref{R1}) (which give solutions by standard binary
Darboux transformation).

\begin{figure}[t]
\begin{minipage}[t]{75mm}\centering
\includegraphics[width=65mm]{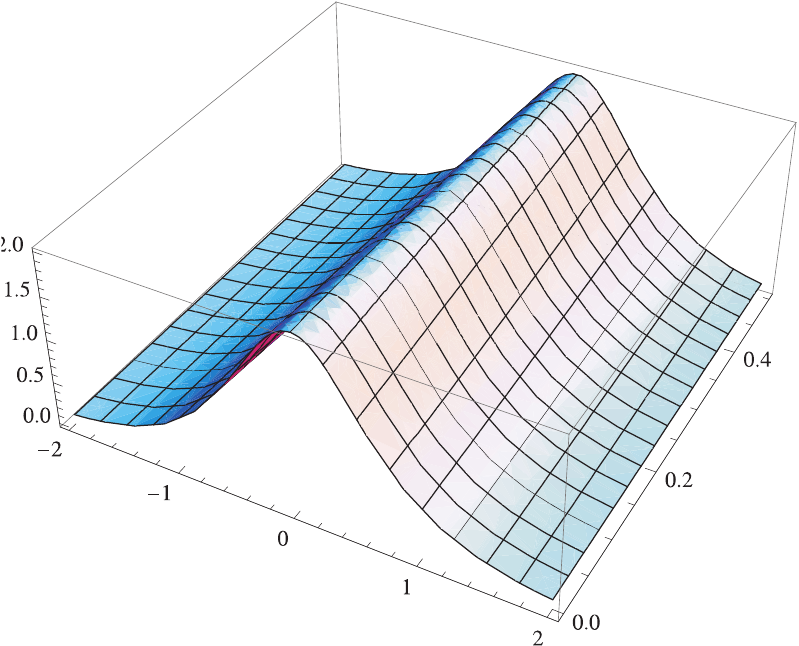}
\caption{Plot of solution~(\ref{R1e}) representing one soliton solution $\partial_x q[1]$.}\label{Fig1}
\end{minipage}\hfill
\begin{minipage}[t]{75mm}\centering
\includegraphics[width=65mm]{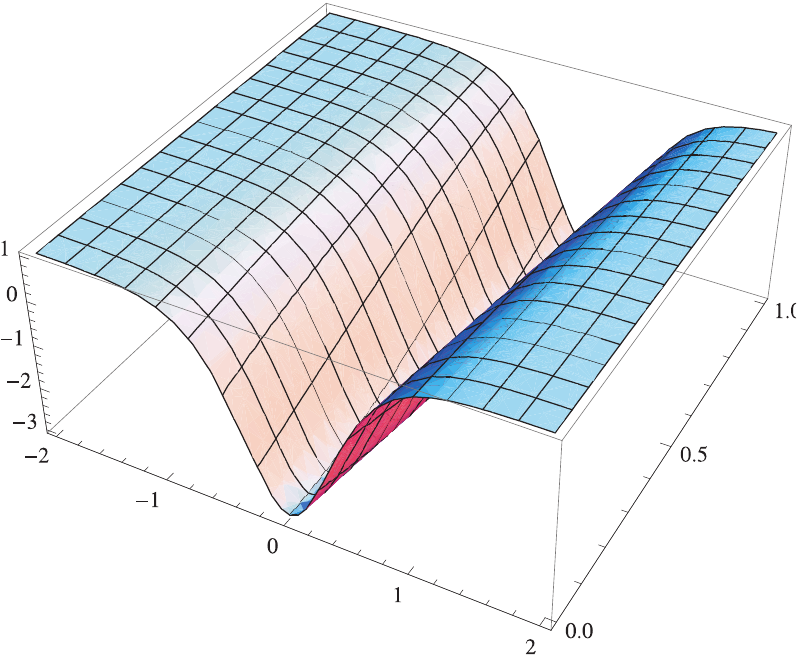}
\caption{Plot of solution (\ref{Q1e}) representing one soliton solution $r[1]$.}\label{Fig2}
\end{minipage}
\end{figure}

It is simple to show that the solutions (\ref{R1e}) and (\ref{Q1e})
have the same behaviour in asymptotic limit and satisfy
(\ref{sprimeneu}). Plots of solutions~(\ref{R1e}) and~(\ref{Q1e})
for $\lambda =i$ are shown in Figs.~\ref{Fig1} and~\ref{Fig2}. Now we show the
relationship between our solution of the system~(\ref{SU2}) and the
solution~$\phi $ of the sine-Gordon equation. The
sine-Gordon equation is given as%
\begin{gather*}
\partial _{x}\partial _{t}\phi =2\sin \phi ,
\end{gather*}
and is related to our system by the following equations%
\begin{gather}
\partial _{x}q  = \cos \phi ,  \qquad 
r  = \pm \frac{1}{2}\partial _{t}\phi .  \label{SGR}
\end{gather}
To obtain the expression for $\phi  [ 1 ] $, we use equation (\ref{SGR}) in equation (\ref{Q1e}) which gives%
\begin{gather}
\pm \frac{1}{2}\partial _{t}\phi  [ 1 ]   = i\lambda
^{-1}\operatorname{sech}\hat{l},  \notag \\
\phi  [ 1 ]   = \pm 2i\lambda ^{-1}\int \operatorname{sech}\left(
2i\lambda x-\frac{i}{\lambda }t\right) dt
 = \pm \frac{2}{\lambda ^{2}}\tan ^{-1}\left( \exp \left( 2i\lambda x-\frac{i}{\lambda }t\right) \right) .  \label{kink}
\end{gather}
The equation (\ref{kink}) is the one-kink solution to the
sine-Gordon equation~\cite{li}.

\section{Conclusions}

In this paper, we
have composed the elementary Darboux transformations of the
generalized coupled dispersionless system and obtained the standard
binary Darboux transformation of the model. By iterating the
standard binary Darboux transformation we have generated the
multisolitons of the model. We have also obtained the
quasideterminant expression for the potential $\Delta $. We have
also considered the case of coupled dispersionless integrable system
based on the Lie group $SU\left( 2\right) $, and have obtained
explicit expressions of Grammian solutions of the system. There are
various directions in which the the integrability properties of the
generalized dispersionless integrable system can be studied. One
such study is to investigate the r-matrix structure and the
existence of inf\/initely many conservation laws of the system. We
shall return these and related investigations in a separate work.

\subsection*{Acknowledgements}

BH would like to thank Department of Physics, University of the
Punjab, Lahore, Pakistan for providing the research facilities.


\pdfbookmark[1]{References}{ref}
\LastPageEnding

\end{document}